\documentclass[10pt,english,a4paper,fleqn,nofootinbib,showpacs,showkeys,aps]{revtex4}
\usepackage[T1]{fontenc}
\usepackage{booktabs}
\usepackage{amsmath}
\usepackage{amssymb}

\makeatletter

\providecommand{\tabularnewline}{\\}

\usepackage{lmodern}
\usepackage{float}
\usepackage{verbatim}

\makeatletter
\newlength{\BCOR}
\newlength{\MARGIN}
\newlength{\TOP}
\newlength{\FOOTABSEP}
\newlength{\FOOTBESEP}
\newlength{\@CHSEP}

\makeatother

\newcommand*{\R}{^3\!{R}}

\DeclareMathOperator{\arcsinh}{arcsinh}

\renewcommand*{\Im}{\textrm{Im}\,}

\makeatother

\usepackage{babel}

\begin{document}

\title{Generalized Painlev\'{e}-Gullstrand metrics}

\preprint{arXiv:0810.2161}

\author{Chun-Yu Lin}

\email{l2891112@mail.ncku.edu.tw}

\author{Chopin Soo}

\email{cpsoo@mail.ncku.edu.tw}

\affiliation{Department of Physics, National Cheng Kung University, Tainan 70101,
Taiwan.}

\begin{abstract}
An obstruction to the implementation of spatially flat Painlev\'{e}-Gullstrand(PG)
slicings is demonstrated, and explicitly discussed for Reissner-Nordstr\"om
and \mbox{Schwarzschild-anti-deSitter} spacetimes. Generalizations
of PG slicings which are not spatially flat but which remain regular
at the horizons are introduced. These metrics can be obtained from
standard spherically symmetric metrics by physical Lorentz boosts.
With these generalized PG metrics, problematic contributions to the
imaginary part of the action in the Parikh-Wilczek derivation of Hawking
radiation due to the obstruction can be avoided.
\end{abstract}

\pacs{04.20.Cv, 04.70.Dy}

\keywords{Painlev\'{e}-Gullstrand metrics, spatial flatness, Hawking radiation,
Parikh-Wilczek method}

\maketitle

\section{Overview}

Painlev\'{e}-Gullstrand(PG) coordinates\cite{PG1,PG2} have often
been employed to study the physics of black holes. They have the
advantage of remaining regular at the horizons, and among the
various regular coordinatizations for spherically symmetric
spacetimes, PG coordinates feature spatially flat slicings with
interesting physical interpretations (see, for instance,
Ref.\cite{Martel:2000rn,Hamilton:2004au}). They have also been
applied to analyse quantum dynamical black
holes\cite{Husain:2005gx}, and used extensively in derivations of
Hawking radiation as tunneling following the work of Parikh and
Wilczek\cite{Parikh:1999mf}. The
existence\cite{Guven:1999hc,Husain:2001wk} and
uniqueness\cite{Beig:2007zz} of spatially flat PG coordinates for
spherically symmetric spacetimes have also been discussed
previously.

In this brief note we demonstrate an obstruction to the
implementation of spatially flat PG slicings. In general the
vierbein fields for different coordinatizations of a metric are
related by local Lorentz transformations, and PG metrics can be
obtained from the standard form of spherically symmetric metrics
(Eq.\eqref{eq:standard_metric} below) by local Lorentz boosts. It
is often taken for granted that spatially flat PG coordinates can
be achieved for arbitrary spherically symmetric spacetimes.
However, the insistence on spatial flatness is an additional
requirement which can lead to complex PG metric variables which
cannot be attained through physical Lorentz boosts. In the next
section, the general condition for real PG metric variables
realized through physical Lorentz boosts is derived. This allows a
general discussion of the obstruction to spatially flat slicings
which we also illustrate with the explicit examples of
Schwarzschild-anti-deSitter(SAdS) and Reissner-Nordstr\"{o}m(RN)
metrics. Spatially flat real PG slicings become problematic at
large distances in the former and at small distances in the
latter. Generalized PG coordinates which remain regular at all
distances are then introduced, and constant curvature slicings are
explicitly constructed and illustrated with
\mbox{Schwarzschild-(anti-)deSitter} spacetimes. A generalized PG
metric which is valid for RN black hole is also constructed
explicitly.

In discussions of black hole evaporation with the Parikh-Wilczek
method\cite{Parikh:1999mf}, PG metrics are often employed because
they are regular at the horizon through which the tunneling
occurs. The tunneling amplitude is associated with the imaginary
contribution of the action which is attributed entirely to the
pole in the integrand. However, the obstruction to spatially flat
PG coordinates can give rise to additional spurious contributions
which are ambiguous and problematic both to the computation of the
tunneling rate and to the universality of the results for
different metrics. We demonstrate that these problems can be
completely avoided by using suitably generalized PG coordinates,
and end with a summary of the explicit results of the tunneling
process for several metrics. We adopt natural units $\hbar=c=1$,
and $k_{B}=1$.

\section{Generalized Painlev\'{e}-Gullstrand coordinates}

We focus on spherically symmetric metrics in four dimensions for
concreteness, although the constructions to be discussed below may
be generalized to higher dimensions. To wit, the most general form
of the metric can be expressed as (see, for instance Sec. 11.7 of
Ref.\cite{WeinbergGC})
\begin{eqnarray}
ds^{2} & = & -N(R,T)dT^{2}+f^{-1}(R,T)dR^{2}+R^{2}d\Omega^{2}.\label{eq:standard_metric}\end{eqnarray}
 To transform to generalized PG coordinates, we introduce the PG time
through%
\footnote{For $dt_P$ time to remain a perfect differential even in
the time-dependent case, an integrating factor should be included
so that $dt_{P} = e^{I(r,T)}\left[dT+\beta(r,T)dr\right]$,
together with the requirement
$\partial_{r}e^I=\partial_{T}(e^I\beta)$ which uniquely specifies
$e^{I(r,T)}$ given its initial value. The metric is still of the
form in \eqref{eq:GPG}, but with $dt_{P}\mapsto e^{-I}dt_{P}$.
This does not effect the expression
of $\beta$ and the rapidity of the Lorentz boost to be discussed later on.%
} $dt_{P}=dT+\beta dr$, and define $L\, g(r)\equiv R(r)$ with real
positive constant length $L$. $dt_{P}=dT+\beta dr$, and define
$L\, g(r)\equiv R(r)$ with real positive constant length $L$. It
follows that\[ ds^{2}=-Ndt_{P}^{2}+2\beta
Ndt_{P}dr+\left(L^{2}f^{-1}g'^{2}-\beta^{2}N\right)dr^{2}+L^{2}g^{2}(r)d\Omega^{2}.\]
 By choosing $\beta=L(Nf)^{\frac{-1}{2}}\sqrt{g'^{2}-pf}$ with $g'(r)\equiv\frac{dg(r)}{dr}$,
the 3-metric on constant-$t_{P}$ hypersurfaces will take the form
$L^{2}\left(pdr^{2}+g^{2}(r)d\Omega^{2}\right)$; and the 4-metric
can thus be written as\begin{eqnarray}
ds^{2} & = & -Nf^{-1}g'^{2}p^{-1}dt_{P}^{2}+p\left(g'^{-1}dR+dt_{P}N^{\frac{1}{2}}f^{\frac{-1}{2}}p^{-1}\sqrt{g'^{2}-pf}\right)^{2}+R^{2}d\Omega^{2}\nonumber \\
 & = & -Nf^{-1}g'^{2}p^{-1}dt_{P}^{2}+p\left(Ldr+dt_{P}N^{\frac{1}{2}}f^{\frac{-1}{2}}p^{-1}\sqrt{g'^{2}-pf}\right)^{2}+L^{2}g^{2}(r)d\Omega^{2}\label{eq:GPG}\end{eqnarray}
 It can be verified that the vierbein fields of this PG metric is
related to those of the standard form \eqref{eq:standard_metric}
by a local Lorentz boost with rapidity $\xi$ given by
$\tanh\xi=g'^{-1}\sqrt{g'^{2}-pf}$ for $f,N\geq0$, and
$\tanh\xi=g'/\sqrt{g'^{2}-pf}$ for $f,N\leq0$. For asymptotically
flat space-times, the constant $p$ parametrizes a family of
foliation wherein $p=\sqrt{1-v_{\infty}^{2}}$ characterizes the
freely falling observer's initial velocity at
infinity\cite{MTW,Martel:2000rn}. In particular with $p=0$ and
$g'=1$ the metric reduces to the usual Eddington-Finkelstein
coordinates%
\footnote{Actually the Kruskal extension of the standard Schwarzschild metric,
covering both black and white holes, must include PG time coordinates
$dT\pm f^{-1}dR$ with both signs, the Eddington-Finkelstein coordinates
$(p=0)$ thus only covers half of the Kruskal extension(see, for instance,
Sec.5.5 of Ref.\cite{HawkingEllis}).%
} :
$ds^{2}=-Ndt_{P}^{2}+2N^{\frac{1}{2}}f^{\frac{-1}{2}}dt_{P}dR+R^{2}d\Omega^{2}$.
Unlike the standard form which has $dT\mapsto-dT$ symmetry, the PG
metric is not invariant under
$dt_{P}\mapsto-dt_{P}$\cite{HawkingEllis}.

It is usually assumed that spherically symmetric metrics can
always be put into the PG form with \textit{spatially flat
slicings} (with $\frac{R}{L}=g(r)=r$ and $p=1$), but our
construction above shows that there is actually an
\textit{obstruction} which prevents this implementation. The
criterion for obtaining real PG metric variables
is\begin{equation} (g'^{2}-pf)\geq0\qquad\forall\,
R.\label{eq:criterion_GPG}\end{equation}
 As mentioned earlier, this is also precisely the condition that the
PG metric can be obtained from the standard form through physical
Lorentz boosts.

To discuss both the obstruction and the extensions to generalized
PG coordinates explicitly, we henceforth consider the case $p=1$
in this section. By letting $\frac{R}{L}=g(r)\equiv r,\sin r\textrm{ or }\sinh r$,
and thus $g'(r)=\sqrt{1-kL^{-2}R^{2}}$ with $k=0,\pm1$, we obtain
constant-$t_{P}$ hypersurfaces which correspond respectively to slicings
with flat, elliptic and hyperbolic 3-geometries. In particular, for
\mbox{Schwarzschild-(anti-)deSitter} metrics, $N=f=1-\frac{2GM}{R}-\frac{\Lambda}{3}R^{2}$;
and the resultant PG metrics are then\begin{equation}
ds^{2}=-\left(1-kL^{-2}R^{2}\right)dt_{P}^{2}+\left[\left(1-kL^{-2}R^{2}\right)^{-1/2}dR+dt_{P}\sqrt{\frac{2GM}{R}+\left(\frac{\Lambda}{3}-\frac{k}{L^{2}}\right)R^{2}}\right]^{2}+R^{2}d\Omega^{2}.\label{eq:constantRPG}\end{equation}
 These have constant curvature 3-metrics $L^{2}\left(dr^{2}+g^{2}(r)d\Omega^{2}\right)$
with Ricci scalar $\R=6k/L^{2}$. The usual spatially flat PG metric
for Schwarzschild black hole coincides with $\Lambda=k=0$, and the
criterion $(g'^{2}-f)=\frac{2GM}{R}\geq0$ clearly holds. However
with non-vanishing, in particular negative, cosmological constant
$\Lambda<0$, we observe that $(g'^{2}-f)=\frac{2GM}{R}+\left(\frac{\Lambda}{3}-\frac{k}{L^{2}}\right)R^{2}<0$
for $R>R_{c}=\sqrt[3]{\frac{6GM}{-\Lambda}}$ if we insist on having
spatially flat $k=0$ slicings. Thus there is an explicit obstruction
which prevents us from achieving spatially flat PG coordinates through
physical Lorentz boosts.

This naturally leads to generalized PG metrics which are not necessarily
spatially flat but remain regular at the horizon(s). PG coordinates
with non-vanishing constant curvature 3-metrics are the obvious candidates
to consider. To wit, our construction above is already adapted to
this analysis. For \mbox{Schwarzschild-(anti-)deSitter} metrics,
the criterion $(g'^{2}-f)=\frac{2GM}{R}+\left(\frac{\Lambda}{3}-\frac{k}{L^{2}}\right)R^{2}\geq0$
can be guaranteed for all $R$ iff $\frac{\Lambda}{3}\geq\frac{k}{L^{2}}$.
This means that for the case of $\Lambda\geq0$ spatially flat $(k=0)$
slicings can be attained, but for the \mbox{anti-deSitter} case,
hyperbolic $(k=-1)$ 3-geometry is needed. Note that for $\Lambda>0$,
all spatial topologies $k=0,\pm1$ are allowed, but for $k=1$ the
range of $R$ is governed by $\frac{R^{2}}{L^{2}}=\sin r\leq1$, yielding
$R\leq L$ which can be as large as possible since $L$ is an arbitrary
parameter which does not affect the 4-metric.

Generalization to PG coordinates beyond constant curvature
slicings may also be required when more generic spherically
symmetric metrics are taken into account. In particular, despite
the attempt to construct flat slicings in \cite{Qadir:2006aj}, the
same obstruction appears near the singularity of the RN metric.
Using our construction above, but now with
$g(r)=\sqrt{r^{2}-L^{-2}O^{2}}$ and
$f=1-\frac{2GM}{R}+\frac{Q^{2}}{R^{2}}$ for RN metric, the
resultant metric from Eq.\eqref{eq:GPG} is then\begin{eqnarray}
ds^{2} & = &
-\left(1+\frac{O^{2}}{R^{2}}\right)dt_{P}^{2}+\left(\frac{R}{\sqrt{R^{2}+O^{2}}}dR+dt_{P}\sqrt{\frac{2GM}{R}+\frac{O^{2}-Q^{2}}{R^{2}}}\right)^{2}+R^{2}d\Omega^{2},\label{eq:GPG_RN}\end{eqnarray}
 wherein $O=0$ corresponds to spatially flat slicings which \textit{fail}
to satisfy the criterion
$(g'^{2}-f)=\frac{2GM}{R}-\frac{Q^{2}}{R^{2}}\geq0$ and give rise
to complex PG coordinates for $R<R_{c}\equiv\frac{Q^{2}}{2GM}$.
This {}``defect'' was considered to be an asset of the PG
formalism \cite{Hamilton:2004au} signaling the presence of
unphysical negative interior mass $M(r)$ within $R_{c}$(see for
instance the discussion in Sec. 31.6 of Ref.\cite{MTW}). But it is
possible to avoid complex PG coordinates if we give up spatially
flat slicings. To wit, we observe that the our criterion
\eqref{eq:criterion_GPG} here is
$\frac{2GM}{R}+\frac{O^{2}-Q^{2}}{R^{2}}\geq0$, which holds if we
choose $O^{2}>Q^{2}$. This implies the constant-$t_{P}$
hypersurfaces (which are regular $\forall R>0$) are no longer
flat; but they can be characterized by the eigenvalues of the
3-dimensional Ricci tensor which are
$\lambda=(0,0,\frac{2O^{2}}{R^{4}})$ indicating that these
hypersurfaces are deviations from flat slicings due to the
parameter $O$. In general, given $f$, an appropriate choice of $g$
can be chosen to obtain the corresponding trouble-free PG
coordinates which however do not always result in spatial
flatness.

\section{Hawking radiation as tunneling\label{sec:Parikh}}

PG metrics, which have the advantage of being regular at the
horizon, have also been employed to study the physics of black
hole evaporation. In their seminal work, Parikh and
Wilczek\cite{Parikh:1999mf} derived Hawking radiation as a
tunneling process. Subsequently the method was generalized to
cases with non-vanishing cosmological
constant\cite{Hemming:2000as,Parikh:2002qh,Medved:2002zj}. In what
follows, we briefly recap the analysis with our generalized PG
coordinates, and illustrate how ambiguities and difficulties which
arise from the obstruction discussed earlier can be avoided by
adopting our generalized PG metrics.

Hawking radiation is treated as tunneling across the horizon from
initial position $R_{i}$ to $R_{f}$ of massless semiclassical
s-wave emission carrying total positive energy $\omega$, and the
black hole with initial mass parameter $M$ shrinks by an amount
$\omega$ in the process thereby maintaining energy conservation.
The metric satisfying Einstein's equations is taken to be
\eqref{eq:GPG} with $N(R)=f(R)$; and the outgoing particles follow
the null geodesic
$\dot{R}\equiv\frac{dR}{dt_{P}}=p^{-1}g'\left(g'-\sqrt{g'^{2}-pf}\right)$.
The decay rate comes from the imaginary part of the particle
action which is associated
with\cite{Parikh:1999mf,Parikh:2002qh}\begin{eqnarray*}
I=\int_{R_{i}}^{R_{f}}p_{R}dR=\int_{R_{i}}^{R_{f}}\left(\int_{0}^{p_{R}}dp_{R}\right)dR
& = &
\int_{R_{i}}^{R_{f}}\int_{H_{0}}^{H_{0}-\omega}\frac{dH}{\dot{R}}dR.\end{eqnarray*}
In the last step, the Hamilton's equation,
$\left.\frac{dH}{dp_{R}}\right|_{R}=\dot{R}$, for the
semiclassical process is invoked. Switching the order of
integration, together with $dH=-d\omega$, yields\begin{eqnarray*}
I & = & \int_{0}^{\omega}\int_{R_{i}}^{R_{f}}\frac{dR}{\dot{R}}(-d\omega')\\
 & = & \int_{0}^{\omega}\int_{R_{i}}^{R_{f}}\frac{g'+\sqrt{g'^{2}-pf}}{g'f}dR(-d\omega'),\end{eqnarray*}
 wherein $f$ in the integrand is evaluated at $M-\omega'$; and the
pole is located at the horizon through which the tunneling occurs
i.e. at $R_{h}$ with $\left.f(R_{h})\right|_{M-\omega'}=0$. The
integral over $R$ is defined by deforming the contour to go through
an infinitesimal semicircle $R=R_{h}+\epsilon e^{i\theta}$ around
the pole%
\footnote{A positive decay rate is associated with clockwise traversal of the
semicircle in the contour\cite{Parikh:1999mf}.%
}, and its imaginary part is then\begin{eqnarray}
\Im\int_{R_{i}}^{R_{f}}dR\frac{g'+\sqrt{g'^{2}-pf}}{g'f} & = & \lim_{\epsilon\rightarrow0}\int_{2\pi}^{\pi}d\theta\epsilon e^{i\theta}\left.\frac{g'+\epsilon e^{i\theta}\partial_{R}g'+\sqrt{g'^{2}-pf+\epsilon e^{i\theta}\partial_{R}(2g'-pf)}}{g'f+\epsilon e^{i\theta}(\partial_{R}g'f+g'\partial_{R}f)}\right|_{R_{h}(\omega')}\label{eq:ImI_pole}\\
 & = & -\left.\frac{2\pi}{\partial_{R}f(R)}\right|_{R_{h}(\omega')},\nonumber \end{eqnarray}
 provided $\sqrt{g'^{2}-pf}$ \textit{remains real} (which is the
same as criterion \eqref{eq:criterion_GPG}). The final result\begin{equation}
\Im I=\int_{0}^{\omega}d\omega'\left.\frac{2\pi}{\partial_{R}f(R)}\right|_{R_{h}(\omega')}\label{eq:ImI}\end{equation}
 governing the decay rate is, remarkably, independent of $g(r)$,
and hence the same for all our generalized PG coordinates
discussed earlier. In Ref.\cite{cmchen} it was shown that $\Im I$
is independent of the coordinates chosen as long as the metric is
regular at the horizon. However, this proof still assumes the
imaginary part of the action can be attributed entirely to the
deformed contour around the pole. For PG coordinates, the result
would be different and problematic due to the spurious
contributions from the failure of \eqref{eq:criterion_GPG} if
spatially flat PG slicings are insisted upon. In particular,
\eqref{eq:ImI_pole} will then acquire an additional spurious
contribution from the imaginary part of the Cauchy principal value
$\Im\left({\mathcal{P}}\int_{R_{i}}^{R_{f}}\frac{\sqrt{1-f}}{f}dR\right)$.
To take the the RN metric as an explicit example: if we define
$\gamma\equiv1-\frac{Q^{2}}{(GM)^{2}}$ with $0\leq\gamma\leq1$
ensuring the existence of an outer horizon, it can seen that
$R_{h}-R_{c}=GM(\sqrt{\gamma}-\gamma)$ can be arbitrarily small
(albeit for rather extreme values of $M$ and $Q$) and thus the
tunneling region will involve values of $R<R_{c}$ for which
spatially flat PG coordinates becomes problematic\footnote{The
case of tunneling through both RN horizons for which expressions
\eqref{eq:ImI_pole} and \eqref{eq:ImI} will receive contributions
from both of the poles has been discussed with
Eddington-Finkelstein ($p=0$) coordinates in Ren J {\it Tunneling
effect of two horizons from a Reissner-Nordstr\"om black hole}
Int. J. Theor. Phys. DOI 10.1007/s10773-008-9818-7.}. Similarly,
with the SAdS metric,
$R_{h}=\frac{2}{\sqrt{-\Lambda}}\sinh\left[\frac{1}{3}\arcsinh\left(3GM\sqrt{-\Lambda}\right)\right]$,
and $R_{c}-R_{h}$ can also be small enough to create difficulties
and ambiguities in the computation of $\Im I$ for spatially flat
PG slicings. These problems can be \textit{avoided altogether} by
adopting our generalized PG metrics which are not spatially flat
in general but chosen to ensure the physical requirement
$\sqrt{g'^{2}-pf}$ remains real.

The change of the Bekenstein-Hawking entropy from $\Delta S=-2\Im
I$ yields, at the lowest order, the temperature from the first law
$T_{\textrm{eff}.}\Delta S=-\omega$ which agrees with the Hawking
temperature\cite{Hawking:1974sw}
$T_{H}\equiv\frac{\kappa}{2\pi}=\frac{1}{4\pi}\left.\partial_{R}f\right|_{R_{h}}$;
but deviations from pure thermal physics indicated by higher order
corrections in $\Im I$ of \eqref{eq:ImI} are displayed in the
table below.

\begin{table}[th]
\begin{tabular}{cccc}
\toprule
\addlinespace
 & Schwarzschild  & Reissner-Nordstr\"om & Schwarzschild (anti-)deSitter\tabularnewline
 & $f=1-\frac{2GM}{R}$ &  $f=1-\frac{2GM}{R}+\frac{Q^{2}}{R^{2}}$ &  $f=1-\frac{2GM}{R}-\frac{\Lambda R^{2}}{3}$\tabularnewline\addlinespace
\midrule
\addlinespace
$2\frac{d\Im I}{d\omega}\qquad$ & $8\pi G\left(M-\omega\right)$ & $\frac{2\pi\left(G(M-\omega)+\sqrt{G^{2}(M-\omega)^{2}-Q^{2}}\right)^{2}}{\sqrt{G^{2}(M-\omega)^{2}-Q^{2}}}$ & $\frac{8\pi\Lambda^{-1/2}\sin\left[\frac{1}{3}\arcsin\left(3G(M-\omega)\sqrt{\Lambda}\right)\right]}{-1+2\cos\left[\frac{2}{3}\arcsin\left(3G(M-\omega)\sqrt{\Lambda}\right)\right]}$\tabularnewline\addlinespace
\bottomrule
\end{tabular}

\caption{In the table, values of $\frac{d\Im I}{d\omega}$ for
various spherically symmetric spacetimes are shown for tunneling
through the outer horizon of RN metric, and through the black hole
horizon for the others. For the SAdS case, the expression
$\sqrt{\Lambda}$ should be understood as $i\sqrt{-\Lambda}$, and
the results are compatible with those for the case of $4+1$
dimensions in Ref.\cite{Hemming:2000as}. For pure Schwarzschild
and RN metrics, the results agree with Ref.\cite{Parikh:1999mf}.
Note also that $\Delta S=-2\Im I$ as evaluated from \eqref{eq:ImI}
coincides with the computation from the area law $\Delta
S=\frac{\Delta A}{4G}=\frac{\pi\Delta R_{h}^{2}}{G}$. }

\end{table}

\vskip-1.0cm
\section*{Acknowledgments}

This work has been supported in part by funds from the National Science
Council of Taiwan under Grant No. NSC95-2112-M-006-011-MY3, and by
the National Center for Theoretical Sciences, Taiwan.

\expandafter\ifx\csname url\endcsname\relax
\newcommand{\url}[1]{\texttt{#1}}
\fi \expandafter\ifx\csname urlprefix\endcsname\relax\def\urlprefix{URL }
\fi \providecommand{\eprint}[2][]{\url{#2}}

\end{document}